\documentclass[prx,nofootinbib,twocolumn,showkeys,superscriptaddress,preprintnumbers,floatfix]{revtex4-1}
\usepackage{appendix}
\usepackage{hyperref}
\usepackage{dynlearn}
\fxsetup{status=draft}
\usepackage{empheq}
\usepackage{overpic}

\makeatletter
\renewcommand\paragraph[1]{%
  \textit{#1.}\hspace{1em}%
}
\makeatother

\newcommand{\kB}{k_\text{B}}

\newcommand{\MSt}{\mathcal{M}}

\newcommand{\drive}{x_{0:\tau}}

\newcommand{\stationary}{\boldsymbol{\pi}}

\newcommand{\zero}{{\color{blue}{\boldsymbol{\mathtt{0}}}}}
\renewcommand{\one}{{\color{blue}{\boldsymbol{\mathtt{1}}}}}  % Note that we do NOT want the `all-ones vector' \one here

\renewcommand{\H}{\operatorname{H}}

\newcommand{\tr}{\text{tr}}

\renewcommand{\TR}{\Theta}
\newcommand{\TRinv}{\TR^{-1}}

\renewcommand{\drive}{\lambda_{0:\tau}}

\makeatletter
\@booleanfalse\titlepage@sw
\makeatother

\begin{document}

\def\ourTitle{Thermodynamic Advantage of Quantum Time-Reversal}%\ourTitle{Time Symmetries of Quantum Memory Improve Thermodynamic Efficiency}

\def\ourAbstract{%
Summary contents of the insightful note.  
}

\def\ourKeywords{%
}
%  nonequilibrium thermodynamics, entropy
%  production, quantum thermodynamics
%}

\hypersetup{
  pdfauthor={Paul M. Riechers},
  pdftitle={\ourTitle},
  pdfsubject={\ourAbstract},
  pdfkeywords={\ourKeywords},
  pdfproducer={},
  pdfcreator={}
}

%%%%%%%%%%%%%%%%%%%%%%%%%%%%%%%%%%%%%%%%%%%%%%%%%%%%%%%%%%%%%%%%%%%%%%%%%%%%%%%

\title{\ourTitle}

\author{Alexander B. Boyd}
\email{alecboy@gmail.com}
\affiliation{Beyond Institute for Theoretical Science (BITS),
San Francisco, CA}

\author{Paul M. Riechers}
\email{pmriechers@gmail.com}
\affiliation{Beyond Institute for Theoretical Science (BITS),
San Francisco, CA}

	%Nanyang Quantum Hub,
	%School of Physical and Mathematical Sciences,
	%Nanyang Technological University, Singapore}

%\author{Chaitanya Gupta}
%\email{something@something}
%
%\affiliation{Nanyang Quantum Hub,
%	School of Physical and Mathematical Sciences,
%	Nanyang Technological University, Singapore}
%
%\author{Artemy Kolchinsky}
%\email{something@something}
%
%\affiliation{Santa Fe Institute}
%
%
%\author{Mile Gu}
%\email{mgu@quantumcomplexity.org}
%
%\affiliation{Nanyang Quantum Hub,
%	School of Physical and Mathematical Sciences,
%Nanyang Technological University, Singapore}

%\affiliation{Unknown details, Caltech}

\date{\today}
\bibliographystyle{unsrt}

\begin{abstract}
Classical computations inherently require energy dissipation that increases significantly as the reliability of the computation improves. 
This dissipation arises when
%This phenomenon arises from irreversibility, where forward 
transitions between memory states are not balanced by their time-reversed counterparts. While classical memories exhibit a discrete set of possible time-reversal symmetries, quantum memory offers a continuum. This continuum enables the design of quantum memories that minimize irreversibility. As a result, quantum memory reduces energy dissipation several orders of magnitude below classical memory.
\end{abstract}

%\keywords{\ourKeywords}

% \pacs{
% 05.45.-a  %  Nonlinear dynamics and nonlinear dynamical systems
% %89.75.Kd  %  Complex Systems: Patterns
% 89.70.+c  %  Information science
% %05.45.Tp  %  Time series analysis
% 02.50.Ey  %  Stochastic processes
% 02.50.-r  %  Probability theory, stochastic processes, and statistics
% 02.50.Ga  %  Markov processes
% 05.20.-y  %  Classical statistical mechanics
% }

%\preprint{\arxiv{1904.XXXX}}

\date{\today}
\maketitle

%\tableofcontents

\setstretch{1.1}

% Collides with table of contents formatting
% \listoffixmes

%\nopagebreak

%\section{Introduction}

\section{Introduction}

Computation requires energy---a fact driven home by Landauer's
recognition that 
``information is physical''~\cite{Land61a}.  
%This is a result of recognizing that ``information is physical.'' 
Memory states of a computation are physically instantiated configurations of a system, meaning they must obey the laws of physics.
For instance, the Second Law of thermodynamics implies that a computation's reduction in state-space must be balanced by entropy flow (e.g., heat) into the environment to guarantee that 
the average entropy production is non-negative. 
 This minimum heat cost, known as Landauer's bound, appears to be inescapable regardless of whether the computation is classical in nature \cite{Land61a,Parr15a} or quantum \cite{Riec21_Impossibility,alicki2004thermodynamics,reeb2014improved}.

Today, practical computation requires much more energy than can be predicted from Landauer's bound \cite{Wims21_Refining, wolpert2024stochastic}.  One might anticipate that this is simply a limitation of our current technology, and that with enough patient engineering we will approach Landauer's bound, achieving the theoretical limit of thermodynamically efficient computing.
However, modern stochastic thermodynamics reveals that 
%common 
realistic
constraints like 
finite duration,
modularity of circuits, 
or a lack of external driving, 
imply fundamentally new bounds on the heat required to drive a computation~\cite{Boyd18a, Riec19_Transforming, wolpert2024stochastic}.
These advances %insights 
have been achieved 
via \emph{equalities} 
known as fluctuation theorems \cite{Croo98a,Jarz00}, which generalize the Second Law inequality, and directly relate entropy production to the dynamics of a computation. 
%Typical assumptions like 
%In the regime of
%overdamped dynamics,

%Under standard assumptions,
Fluctuation theorems imply that a system's observed trajectories bound entropy production. 
 The entropy production quantifies thermodynamic irreversibility and thus loss of energetic resources beyond Landauer's bound~\cite{cisneros2023dissipative, Rold10a, roldan2012entropy, skinner2021improved}.  The 
relative probabilities 
of transitioning between two physical states
%$\frac{p(s) p(s \to s')}{p(s') p(s' \to s)}$ (i.e., the ratio of joint probabilities of
%transitioning from physical states $s$ to $s'$ compared to 
%transitioning from $s'$ to $s$) 
can only be biased via an irreversible investment in thermodynamic resources.
In this vein, 
%the trajectories 
Ref.~\cite{Riec20_Balancing}
showed that
unreciprocated transitions
between %coarse-grained 
memory states of a computer imply 
%(logarithmically) 
divergent energetic costs
as the computer becomes more reliable.  However, 
%in general, thermodynamic dissipation does not depend solely on the relative probabilities of observed transitions.
%This 
this glosses over a nuance that can have profound implications:
\emph{Time reversal symmetries of the physical states co-determine
thermodynamic irreversibility of an observed trajectory}.
While this dependence on time-reversal symmetries is known in nonequilibrium thermodynamics~\cite{jarzynski2004classical}, the implications for the energetics of computation remain largely unexplored besides Refs.~\cite{Riec20_Balancing, boyd2021time}.
Indeed,
the thermodynamic irreversibility of a computation depends sensitively on the 
time-reversal symmetries of memory elements used in the computation.
And, as we will show, quantum can help.

%though---a nuance which is often glossed over.

%However, modern stochastic thermodynamics generalizes %reframes 
%the Second Law 
%via %as 
%\emph{equalities} known as fluctuation theorems \cite{Croo98a,Jarz00}, which directly relate entropy production to the dynamics of a computation.  Specifically, any [thermodynamically?] \emph{irreversible dynamics} in a system comes at an entropic cost \cite{cisneros2023dissipative, Rold10a, roldan2012entropy, skinner2021improved} [PMR: What are we really saying here? 
 %As written, it's a tautology.].  
 %As a result,  irreversible computations require energy dissipation beyond Landauer's bound, which diverges with the reliability of the computation \cite{Riec20_Balancing, boyd2021time} [PMR: we should be really explicit that this assumes time-symmetric driving of metastable memory elements].  Irreversibility appears to be a fundamental resources, which cannot be injected into a computation without a thermodynamic cost.

%The thermodynamic irreversibility of a dynamical process depends sensitively on the type of system that encodes the computation.  

%Time reversal symmetries of the memory describe how the information bearing degrees of freedom 
%change
%when time is reversed.  

For time-even variables like DNA nucleotide sequences, time reversal doesn't affect the encoded information (thymine remains thymine in reverse time). However, information in a hard disk drive is encoded in magnetic moments, which flip sign under time reversal, mapping 
the logical elements
$\zero \mapsto \one$ and $\one \mapsto \zero$.  
Memory devices which change under time reversal, like magnetic storage, are \emph{time-odd} \cite{spinney2012nonequilibrium,proesmans2019hysteretic}.  The flexibility of time-symmetries of memory appears to allow the designer of a computational circuit to choose an appropriate substrate, that effectively minimizes the 
thermodynamic irreversibility of the computation~\cite{boyd2021time}.
%[irreversibility of the computation and associated entropy production] 
%\cite{boyd2021time}.

However, even with the flexibility to design time-symmetries of classical memory, logical irreversibility requires divergent dissipation
in the limit of increasing reliability ~\cite{boyd2021time}. %The standard paradigm for computation requires metastable memories and time-symmetric driving. 
%of time-symmetric protocols (e.g., standard computer clocking)
%acting on metastable memories~\cite{boyd2021time}.  
%This is where 
It is in this context that
we show a considerable energetic advantage 
enabled by quantum memory elements.
%to quantum computation.  
Unlike classical time-reversal, which is limited to discrete involutions, quantum time-reversal operations correspond to anti-unitary operators \cite{Wigner60_Normal, roberts2017three,sakurai2020modern}. In the following, 
we show that the additional flexibility of quantum memory allows us to implement logically irreversible computations without divergent dissipation.  Thus, there is a significant thermodynamic advantage to performing fundamental logical operations with quantum memory.  We show that quantum time-reversal symmetry changes the scaling of dissipation with reliability.

\section{Thermodynamic Bound on Computation}

\emph{Irreversibility of any form is dissipation.}  Detailed fluctuation theorems \cite{Croo99a,Jarz00,esposito2009nonequilibrium}  relate the probability of a stochastic system  trajectory $s_{0:\tau}$ and its reverse trajectory $s_{0:\tau}^R$ to the entropy production.  Specifically, the entropy production of a system that is driven along the trajectory $s_{0:\tau}$ by a control parameter protocol $\lambda_{0:\tau}$ is
\begin{align}
    \Sigma(s_{0:\tau}|\lambda_{0:\tau})=k_B \ln \frac{\Pr(s_{0:\tau}|\lambda_{0:\tau})}{\Pr(s^R_{0:\tau}|\lambda^R_{0:\tau})},
\end{align}
where $\Pr(s^R_{0:\tau}|\lambda^R_{0:\tau})$ gives the hypothetical probability of the reverse trajectory if a reverse experiment were implemented.

Consider a computation, which is instantiated within a memory-storing system $\mathbf{S}$ 
with reduced density matrix $\rho_t$ at time $t$
in contact with a thermal environment $\mathbf{E}$ 
with reduced density matrix $\rho_t^\text{env}$
that starts in local
equilibrium $\rho_0^\text{env} = \bigotimes_b \stationary_b$,
where $\stationary_b$ is the equilibrium distribution (e.g., canonical or grand canonical) for environmental bath $b$.
%$ = \stationary$.  
The computation is implemented through a unitary $U_{\lambda_{0:\tau}}$ applied to the joint system--environment supersystem, where $\lambda_{0:\tau} \equiv \lambda_0 \cdots \lambda_\tau$ is the control sequence.  This is the unitary dynamics that results from applying Hamiltonian $H_{\lambda_t}$ using control parameter $\lambda_t$ at every time $t$ within the computation interval $t \in [0,\tau]$.    The environment adds additional degrees of freedom such that the operation over the system $\mathbf{S}$ need not be unitary and information preserving.  Generally, the net operation over the system $\mathbf{S}$ can be expressed as a completely positive and trace preserving (CPTP) operation $O_{\lambda_{0:\tau}}$.  
This is a 
general framework for quantum open-system transformations
%, and takes 
%the perspective of the ``church of the larger Hilbert space,'' in which 
since, via Stinespring's dilation theorem, 
every CPTP operation on the system can be %interpretted as 
implemented through
a unitary transformation on a larger space~\cite{stinespring1955positive, binder2016work, wilde2013quantum}. 

On the joint system--environment supersystem,
Ref. \ref{app:Microscopic Reversibility} derives a statement of microscopic reversibility:
\begin{align}
\Pr(z \stackrel{\lambda_{0:\tau}}{\mapsto} z') = 
\Pr(z'^\dagger \stackrel{\lambda_{0:\tau}^R}{\mapsto} z^\dagger),
%
%\Pr(Z_\tau=z'|Z_0=z,\Lambda_{0:\tau}=\lambda_{0:\tau})
%=
%\Pr(Z_\tau=z^\dagger|Z_0=z'^\dagger,\Lambda_{0:\tau}=\lambda^R_{0:\tau})
\label{eq:MicroRev}
\end{align}
which, in turn,
allows us to derive thermodynamic consequences for computational transitions on the memory subsystem.
Eq.~\eqref{eq:MicroRev}
says that the probability that (i) a measurement of the joint state would yield $z'$ at the end of a protocol $\lambda_{0:\tau}$ (given 
the initial quantum state $z$)
%that the joint state started as $z$) 
is the same as (ii) the probability of obtaining the measurement outcome
$z^\dagger$---the time-reversal of the joint state $z$---at the end of the time-reversed protocol
$\lambda^R_{0:\tau}$ 
if the supersystem begins in the time-reversed state $z'^\dagger$.

We connect the dynamics of the computation $O_{\lambda_{0:\tau}}$ to thermodynamics via a quantum detailed fluctuation theorem (QDFT) \cite{campisi2011colloquium, jarzynski2004classical,landi2021irreversible}. App. \ref{app:DFT}
derives a general QDFT that directly relates entropy production $\Sigma$ in a quantum transformation to the forward and reverse measurement probabilities on the system alone.  This derivation relies on a two-point measurement framework \cite{esposito2009nonequilibrium, aguilar2022two}.

The generality of this theorem is tied to the fact that it accommodates a system in contact with arbitrarily many baths that begin in local equilibrium, like heat and particle baths that each start as a grand canonical ensemble.  As such, the first contribution to the total entropy production of a system $\mathbf{S}$ is the entropy flow $\Phi$, 
%which we use to refer the entropy changes 
due to particle and heat flow into the environment,
\begin{align}
\Phi = \sum_{b}\frac{Q_b-\mu_b \Delta N_b}{T_b},
\end{align}
where each reservoir is indexed by $b$ with chemical potential $\mu_b$ and temperature $T_b$ \cite{riechers2021initial}.  
This admits a wide variety of possible scenarios for nonequilibrium computation, and follows roughly the same logic as Jarzynski's original Hamiltonian derivation of a detailed fluctuation theorem \cite{Jarz00}.

The total entropy production also includes the change in system 
surprisal
$\Delta S_{s_0,s_\tau} = \kB  \ln \braket{s_0 |\rho_0 |s_0 } - \kB  \ln \braket{ s_\tau | \rho_\tau | s_\tau}$,
%entropy $\Delta S = \kB \tr(\rho_0 \ln \rho_0) - \kB \tr(\rho_\tau \ln \rho_\tau)$,
where $\kB$ is Boltzmann's constant.  $\Sigma_{s_0,s_\tau} \equiv \Phi_{s_0,s_\tau}+\Delta S_{s_0,s_\tau}$ captures the entropy production of a particular measurement trajectory $s_0 \rightarrow s_\tau$, because it produces the change in von Neumann entropy of the system plus the entropy flow to the environment when averaged over all state sequences \cite{riechers2021initial,deffner2011nonequilibrium,Espo10a}.  Note that this entropy production is specifically from the initial post-measurement state $\sum_{s_0} \Pi_{s_0}\rho_0 \Pi_{s_0}$ to the final post-measurement state $\sum_{s_\tau} \Pi_{s_\tau}\rho_\tau \Pi_{s_\tau}$. As Ref. \ref{app:DFT} shows, the entropy production for a state transition $s_0 \rightarrow s_\tau$ is bounded below by \emph{computational entropy production}
%\begin{align}
%\Sigma(s_0,s_\tau) \leq \Sigma = 
%\Phi+\Delta S_\mathbf{S} ,
%\end{align}
\begin{align}
\Sigma^\text{comp}_{s_0,s_\tau} \leq \Sigma_{s_0,s_\tau}.
\end{align}
The subscript $s_0,s_\tau$ indicates the average of all measured trajectories for which the system starts in $s_0$ and ends in $s_\tau$.  This measure quantifies the irreversibility of a state transition
\begin{align}
\label{eq:CompEntProd}
    \Sigma^\text{comp}_{s_0,s_\tau}  \equiv \kB \ln \frac{f(s_0,s_\tau)}{r(s_0,s_\tau)},
\end{align}
in terms of the forward transition probability $f(s_0,s_\tau)$ and reverse transition probability $r(s_0,s_\tau)$.  
%
%Any observed irreversibility in a system requires 
Observed irreversibility in a system typically implies
dissipation \cite{roldan2021quantifying, martinez2019inferring, harunari2022learn}, and $\Sigma^\text{comp}_{s_0,s_\tau} $ is the unavoidable entropy production associated with the computation when $s_0 \rightarrow s_\tau$.  

Crucially in evaluating irreversibility, the time-reversed transition probability $r(s_0,s_\tau)$ involves not only the time-reversed state sequence, but also the physical states $s_0^\dagger$ and $s_\tau^\dagger$ induced by the time-reversal operator acting on 
states $s_0$ and $s_\tau$.
%of the physical states.
Specifically, the forward probability $f(s_0,s_\tau)$ is that of measuring $s_0$ then $s_\tau$ after applying the forward control $\lambda_{0:\tau}$.  The reverse probability $r(s_0,s_\tau)$
is that of measuring $s_\tau^\dagger$ then $s_0^\dagger$ after applying the reverse 
control sequence
$\lambda^R_{0:\tau}$
%$\lambda^R_{0:\tau}=\lambda_{\tau}^\dagger \cdots \lambda_0^\dagger$,  
(where $\lambda_t^R$ applies the time-reversal of %the Hamiltonian 
$H_{\lambda_{\tau-t}}$), assuming that the initial state density is the same as the time reversal of the final state of the forward experiment \cite{Croo99a}.

We can exactly calculate the computational entropy production for a quantum computation using Eq.~\eqref{eq:CompEntProd}.  Given the initial density $\rho_0$, the forward computation $O_{\lambda_{0:\tau}}$, the final density $\rho_\tau = O_{\lambda_{0:\tau}}(\sum_{s_0} \Pi_{s_0} \rho_0  \Pi_{s_0})$, and the reverse computation $O_{\lambda^R_{0:\tau}}$, we can exactly calculate the transition probabilities:
\begin{align}
    f(s_0,s_\tau) & =\text{Tr}\left[ \Pi_{s_\tau} O_{\lambda_{0:\tau}} ( \Pi_{s_0}) \right]p_0(s_0)
   \\  r(s_0,s_\tau) & =\text{Tr}\left[ \Pi_{s_0^\dagger} O_{\lambda^R_{0:\tau}} ( \Pi_{s_\tau^\dagger}) \right]p_\tau(s_\tau).
\end{align}
Here, $\Pi_s \equiv |s \rangle \langle s|$ is the projector onto the pure system state $|s \rangle$, $p_0(s_0) \equiv \text{Tr}\left[ \Pi_{s_0} \rho_0 \right]$ is the initial probability of $s_0$, and $p_\tau(s_\tau) \equiv \text{Tr}\left[ \Pi_{s_\tau} \rho_\tau \right]$ is the final probability of $s_\tau$.  The time reversal of a state comes from the anti-unitary time-reversal operator 
\begin{align}
    |s^\dagger \rangle \equiv \TR |s \rangle.
\end{align}
The way in which the states change under time reversal directly affects the lower bound on entropy production.

\section{Time-Symmetric Control of Classical Computation}

Practical computing paradigms operate under time-symmetric control \cite{Riec20_Balancing,ray2023thermodynamic}.  This restriction is also assumed for nonequilibrium steady state biochemical computing \cite{roldan2010estimating, martinez2019inferring}, thermodynamic clocks \cite{pearson2021measuring}, thermodynamic uncertainty relations \cite{barato2015thermodynamic, hasegawa2019fluctuation}, and thermodynamic cost of waiting time distributions \cite{skinner2021estimating}.  Thus, the ``restriction'' that we make explicit here is in fact a latent assumption pervading much of modern nonequilibrium thermodynamics.  The near-universal adoption of time-symmetric control reflects its practical utility but also obscures its conceptual role: without it, dissipation can be made arbitrarily small by sufficiently slow control \cite{boyd2022thermodynamic}

Time symmetric-control means that the control trajectory is the same under time reversal $\lambda_{0:\tau}=\lambda_{0:\tau}^R$. 
This is most clear for something like a laptop computer which runs its program through the constant nonequilibrium driving of the battery. This remains true even if we consider the computational memory system driven by the periodic $\sim$3GHz clock that marches the Markovian memory dynamics forward.  
For quantum computers too, we can expect a constant or periodic influence to march the computation forward,
whereas a time-asymmetric driving signal would come with its own thermodynamic cost.  Our argument that irreversibility of any form dissipates thermal resources, even in the control protocol $\lambda_{0:\tau}$, suggests that the ubiquity of time-symmetric control in computing is not an accident. Rather, it is a means of avoiding dissipation.  For a deterministically applied protocol, the only way to avoid divergent dissipation is to ensure that it is the same in forward and reverse.

In the case of time-symmetric control, we can identify a single CPTP operation for forward and reverse $O=O_{\lambda_{0:\tau}}=O_{\lambda^R_{0:\tau}}$, which determines the computational entropy production of a particular state sequence
\begin{align}
    \Sigma^\text{comp}_{s_0,s_\tau} = \kB \ln \frac{\text{Tr}\left[ \Pi_{s_\tau} O ( \Pi_{s_0}) \right]p_0(s_0)}{\text{Tr}\left[ \Pi_{s_0^\dagger} O ( \Pi_{s_\tau^\dagger}) \right]p_\tau(s_\tau)} ~.
\end{align}

To appreciate the flexibility of the time-symmetries of quantum mechanics, consider 
first a classical computation that is specified by a Markov channel $M_{s_0 \rightarrow s_\tau} \equiv \Pr(S_\tau=s_\tau|S_0=s_0)$,
which determines the probability of final state $s_\tau$ given the initial state $s_0$, with $S_t$ denoting the random variable for the system at time $t$ \cite{Cove91a}.  We can implement this classical computation as a CPTP map 
\begin{align}
O(\circ) = \sum_{s'_0,s'_\tau}M_{s'_0 \rightarrow s'_\tau} \Pi_{s'_\tau} \text{Tr} \left[ \Pi_{s'_0} \circ \right].
\end{align}
This operation measures in the original computational basis, updates according to the stochastic channel $M$, then outputs a state in the same computational basis.  
 
The resulting 
%expression for 
computational entropy production
\begin{align}
\Sigma^\text{comp}_{s_0,s_\tau} & = \kB \ln \frac{M_{s_0 \rightarrow s_\tau}p_0(s_0)}{M^\dagger_{s_\tau^\dagger \rightarrow s_0^\dagger} p_\tau(s_\tau)} ~,
\end{align}
with the transpose of the reverse computation matrix in the denominator %is defined
\begin{align}
M^\dagger_{s_0^\dagger \rightarrow s_\tau^\dagger} 
 \equiv \sum_{s'_0,s'_\tau}p(s^\dagger_\tau|s_\tau') M_{s'_0 \rightarrow s'_\tau}p(s_0'|s^\dagger_0) ~,
\end{align}
depends on the overlap between states and their time reversals. 
$p(s^\dagger|s')=p(s'|s^\dagger)=|\langle s'|s^\dagger \rangle|^2$ is both the probability of measuring the time reversed state $|s^\dagger \rangle $ given the state $|s'\rangle$, and the probability of measuring $|s' \rangle $ given the time reversed state $|s^\dagger \rangle$.   $M^\dagger_{s_\tau^\dagger \rightarrow s_0^\dagger}$ is the probability of measuring $|s_0^\dagger \rangle$ given that the system is prepared in state $|s_\tau^\dagger \rangle$, measured in the forward-time computational basis, then transformed by the Markov channel $M$.  From an operator perspective, we can express the linear operator for the reverse computation
\begin{align}
    \hat{M}^\dagger= \hat{R}\hat{M}\hat{R}^T,
\end{align}
where
\begin{align*}
    \hat{M} & \equiv \sum_{s,s'}M_{s \rightarrow s'}|s' \rangle \langle s|
    \\ \hat{M}^\dagger & \equiv \sum_{s,s'}M^\dagger_{s^\dagger \rightarrow s'^\dagger}|s' \rangle \langle s|
    \\ \hat{R} & \equiv \sum_{s,s'}p(s'^\dagger|s)|s' \rangle \langle s|.
\end{align*}
Using the reverse computation, we evaluate the average computational entropy production over all input-output combinations $s_0$ and $s_\tau$ to evaluate the minimum thermodynamic efficiency of a computation
\begin{align}
\label{Eq:EntrProd}
\langle \Sigma^\text{comp} \rangle  = \sum_{s_0,s_\tau}\Pr(S_0=s_0,S_\tau=s_\tau) 
\Sigma^\text{comp}_{s_0,s_\tau}
~.
\end{align}

\section{Time-Reversal Operators}

There are many different time-reversal symmetries depending on how information is physically encoded.  In classical physics, these symmetries correspond to involutions, which are functions between elements of the state space $\Theta: \mathcal{S} \rightarrow \mathcal{S}$ that are their own inverse $\Theta(\Theta(s))=s$.  This guarantees that reversing the time twice takes you back to where you started.  Two fundamental examples are time-even positional and time-odd momentum time-reversal, 
\begin{align}
\TR (x)=x \text{, }\TR (p)=-p,
\end{align}
reflecting that positions are preserved while momentum flips sign under time reversal.  
%It is worth noting that these two different types of memory are conjugates of each other.
%PMR: Why is it worth noting?

Quantum time-reversals are much more flexible than classical, because they are antiunitary operators \cite{Wigner60_Normal, roberts2017three, sakurai2020modern}.  Such operators are \emph{antilinear}, meaning that they conjugate when they commute with complex numbers $\TR c = c^* \TR$.  In essence, this means that it can be expressed as the composition of a 
basis-dependent 
unitary $U$ and a complex conjugation operator $K$ 
\begin{align}
    \TR = U K.
\end{align} 
Unlike linear operators, the action of the complex conjugation operator depends on the particular choice of basis in which you choose to represent the memory states. It conjugates coefficients $Kc=c^*K$, but leaves basis elements fixed: 
$K |s \rangle=|s \rangle$ \cite{sigwarth2022time}.  We use this to our advantage in the next section, where we introduce energy-efficient computations with quantum memory.

\section{Quantum Thermodynamic Advantage}

\begin{figure}
\includegraphics[width=\columnwidth]{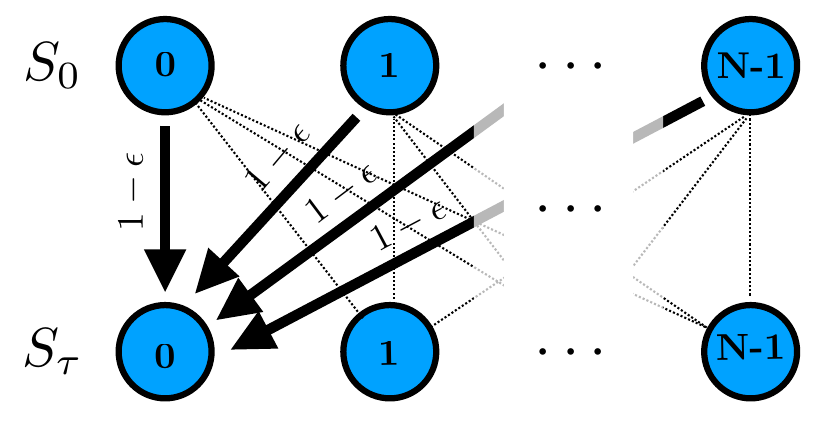}
\caption{\textbf{$N$-state erasure with error $\epsilon$} maps every state to the default state ($0$ in this case) with probability $1-\epsilon$ (highlighted with thick arrows). Every other transition (indicated with thin dashed lines) has probability $\epsilon/(|\mathcal{S}|-1)$.}
\label{fig:N_Erasure} 
\end{figure}

Quantum memory offers a divergent thermodynamic advantage for logically irreversible computation.  Consider the most basic irreversible computation: erasure.  The erasure resets a binary memory $\mathcal{S}=\{ \zero, \one \}$ to a single state, which we will choose to be $\zero$ without loss of generality.   The components of such an operation are 
\begin{align}
\label{eq:Erasure}
M^\text{erase}_{s \rightarrow s'} = \delta_{0,s'}(1-\epsilon) +(1-\delta_{0,s'})\frac{\epsilon}{|\mathcal{S}|-1},
\end{align} 
where $\epsilon \ll 1$ specifies the error rate and $|\mathcal{S}|$ is the dimension of the system.  Note that the expression in Eq. \ref{eq:Erasure} generalizes to an erasure operation that resets an $N$-state memory to a default state, as shown in Fig. \ref{fig:N_Erasure}.  We begin by considering \emph{binary} erasure.

In the special case where memory is classical, we have only two time-reversal symmetries for binary information storage:
\begin{enumerate}
\item Time-even $\zero^\dagger=\zero$ and $\one^\dagger=\one$.
\item Time-odd $\zero^\dagger=\one$ and $\one^\dagger=\zero$.
\end{enumerate}
In the time-even case, the lower bound on entropy production found in Eq.~\eqref{Eq:EntrProd} is
\begin{align}
    \frac{\langle \Sigma^\text{comp}
    \rangle}{k_B}  = & \H[S_\tau]-\H[S_0]+(p_0(\one)-\epsilon ) \ln \frac{1-\epsilon}{\epsilon} ~.
\end{align}
In the time-odd case, the same computational entropy production is
\begin{align}
    \frac{\langle \Sigma^\text{comp} \rangle}{k_B}  =\H[S_\tau]-\H[S_0]+(p_0(\zero)-\epsilon ) \ln \frac{1-\epsilon}{\epsilon}.
\end{align}
In either case, as the reliability increases ($\epsilon \rightarrow 0$), entropy production diverges logarithmically as $\ln(1/\epsilon)$.  As a result, classically implemented erasure must dissipate divergent energy as deterministic computations approach perfect fidelity \cite{Riec20_Balancing}.

By contrast, let's consider the entropy production of quantum positional information bearing degrees of freedom, which are invariant under time reversal \cite{sigwarth2022time}
\begin{align}
    \Theta |x_0 \rangle  = |x_0 \rangle \text{, } \; \Theta |x_1 \rangle & = |x_1 \rangle.
\end{align}
This is in line with quantum computing protocols that store information spatially, for instance in metastable double-wells \cite{riechers2024thermodynamically}.  In this basis, $\Theta$ operates as the complex conjugate operator $\Theta=K$.  However, unitary shifts of the computational basis change the time-reversal.  For instance, if we choose a new computational basis $\{ |a \rangle , |b \rangle \}$ for the qubit which is a superposition of the positional states
\begin{align}
    |a \rangle \equiv \frac{|x_0 \rangle + i |x_1 \rangle}{\sqrt{2}} \text{, } \; 
    |b \rangle \equiv \frac{|x_0 \rangle - i |x_1 \rangle}{\sqrt{2}}
\end{align}
our memory becomes time-odd: $\Theta |a \rangle = |b \rangle$, $\Theta|b \rangle = |a \rangle$.  Surprisingly, both classical binary time-reversal symmetries can be realized in the same system by simply changing the basis of measurement.  This perhaps makes sense, considering that position $\Theta |x \rangle =|x \rangle$ and momentum $\Theta |p \rangle = |- p \rangle$, the canonical examples of time-even and time-odd variables, are conjugate bases occupying the same Hilbert space.

\begin{figure}
\includegraphics[width=\columnwidth]{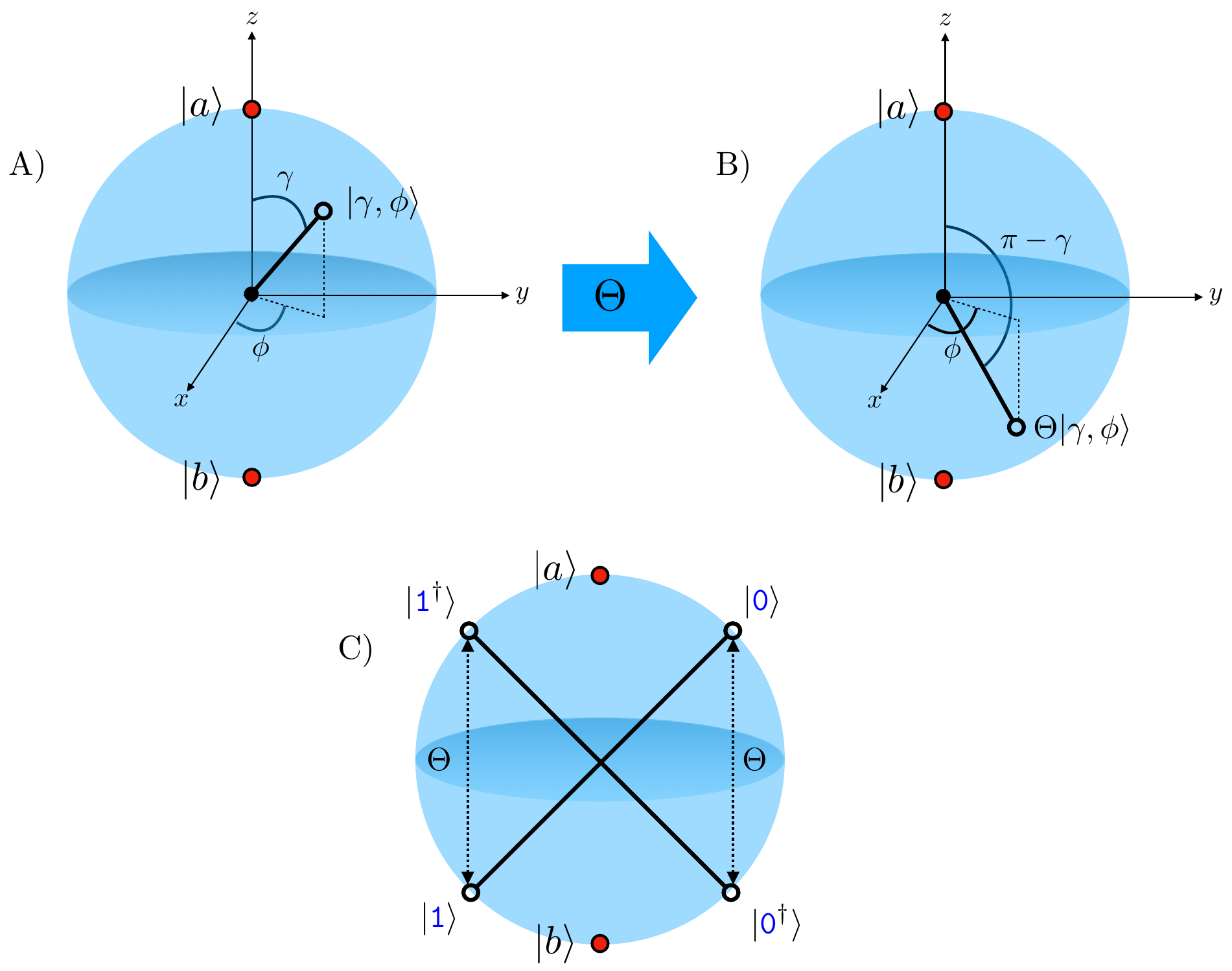}
\caption{Time reversal on the Bloch sphere is a reflection.  The state $|\gamma, \varphi \rangle$ shown in A) is reflected across the $xy$-plane to produce the time reversal in $\Theta |\gamma, \varphi \rangle=|\pi-\gamma, \varphi \rangle$ in B).  C) shows a mutually unbiased basis in the Bloch sphere, where the time reversal $|0^\dagger \rangle$ has the same overlap with $|0\rangle$ as $|1 \rangle$.}
\label{fig:BlochReversal} 
\end{figure}

Unlike classical memory, the continuum of pure quantum states provides a continuum of possible time-reversal symmetries.  As shown in App. \ref{app: Binary Quantum Bases},  if we parametrize the states in the Bloch sphere with the angle $\gamma$ from $|a \rangle$ and the azimuth $\phi$
\begin{align}
    | \gamma ,\phi \rangle = \cos ( \gamma/2) |a \rangle + e^{i \phi} \sin (\gamma/2) |b \rangle,
\end{align}
then time reversal reflects the state across the plane of $\gamma= \pi/2$
\begin{align}
    \Theta |\gamma, \phi  \rangle = e^{- i \phi}| \pi - \gamma , \phi \rangle.
\end{align}
Fig. \ref{fig:BlochReversal} shows how the time reversal applies this reflection in the Bloch sphere.  We identify a particularly efficient informational basis:
\begin{align}
| \zero \rangle  = |\pi/4, \phi \rangle, | \one \rangle  =  | \pi+ \pi/4,\phi \rangle.
\end{align} 
The time-reversal of this quantum basis behaves very differently from classical memory, providing a radical improvement in thermodynamic efficiency.

\begin{figure*}
\includegraphics[width=2\columnwidth]{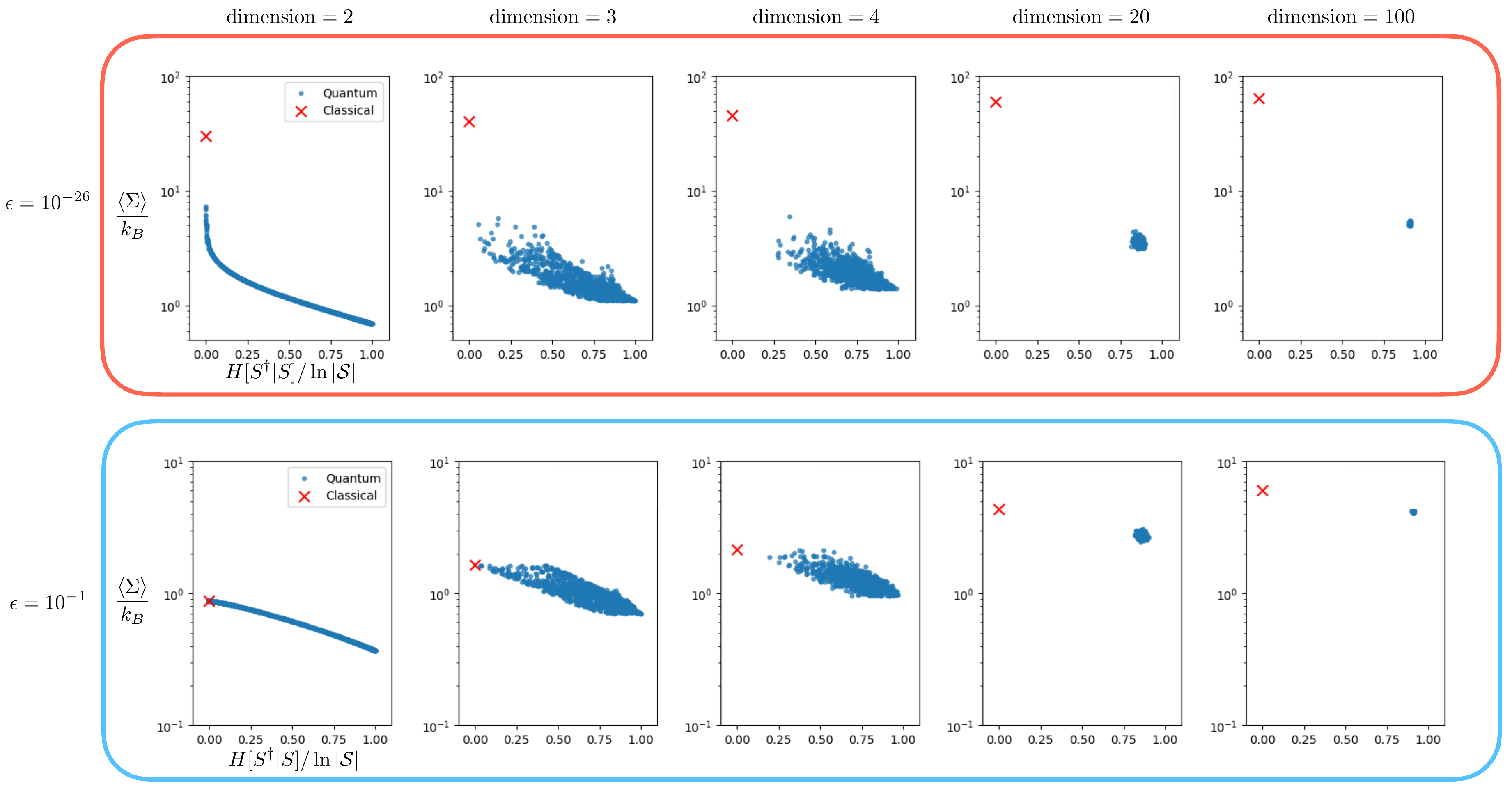}
\caption{\emph{Entropy production decreases as QTR ambiguity increases}. For two different error rates $\epsilon \in \{10^{-1},10^{-26}\}$ and five different system dimensions $|\mathcal{S}| \in \{2,3,4,20,100\}$, we plot the entropy production and QTR ambiguity of an erasure operation for 1) a classical memory (a red X), and 2) one thousand randomly sampled quantum memory bases (blue dots).  Note that the QTR ambiguity $H[S^\dagger|S]$ is scaled by its maximum value $\ln |\mathcal{S}|$.}
\label{fig:Haar_Dissipation} 
\end{figure*}

In calculating the entropy production associated with an erasure in the $\{|\zero \rangle, |\one \rangle \}$ basis above, we see that overlap between each combination of $|s^\dagger \rangle $ and $|s' \rangle $ is the same for all state pairs:
\begin{align}
p(s^\dagger|s')=|\langle s' | s^\dagger \rangle|^2 & = 1 / |\mathcal{S}| %\frac{1}{2}
~.
\end{align}
The memory basis and its time reversal are mutually unbiased \cite{durt2010mutually}.  In this case, and more broadly for a system of any size where the time-reversal is mutually unbiased $|\langle s' | s^\dagger \rangle |^2  = \frac{1}{|\mathcal{S}|}$ the entropy production simplifies to
\begin{align}
    \Sigma^\text{comp}_{s_0,s_\tau} = \kB \ln |\mathcal{S}| + \kB \ln \Pr(S_0=s_0|S_\tau=s_\tau).
\end{align}
The resulting average computational entropy production is
\begin{align}
\label{eq:LowEnt}
\langle \Sigma^\text{comp} \rangle = \kB\ln |\mathcal{S}|- \kB \H[S_0|S_\tau],
\end{align}
where $\H[S_0|S_\tau]$ is the Shannon entropy of the initial state conditioned on the final state (see App. \ref{app:Random Variables}). This entropy production never diverges, unlike computation instantiated on classical memory, whose dissipation necessarily diverges for reliable logically irreversible transformations \cite{boyd2021time}.  In fact, for the specific case of erasure, the uncertainty in the initial state given the final state is maximal $H[S_0|S_\tau]= \log |\mathcal{S}|$, meaning that the \emph{entropy production is zero}.  Therefore, by storing information in quantum mutually unbiased bases, we free low-error irreversible computation from the constraint of divergent dissipation (on the order of $\ln (1/\epsilon)$), and unlock the possibility of \emph{perfectly efficient computing}.  This indicates a thermodynamic advantage to quantum hardware even when implementing classical algorithms.

\section{Exploring Efficiency of Time Symmetries} 

The previous section clearly demonstrates a quantum advantage for a particular designed basis.  We now examine the properties of quantum time-reversal that improve thermodynamic efficiency, which are rooted in the distribution of time-reversed states.  The probability of the time-reversed state given the initial state $p(s^\dagger|s')=|\langle s^\dagger|s' \rangle|^2$ is the source of the quantum thermodynamic advantage for irreversible classical computations.  Unlike classical time reversals, for which the probability is highly peaked around a particular involution $p(s^\dagger|s')=\delta_{s,f(s')}$, quantum time reversals can have high ambiguity in the time-reversed state.  We saw this above, where the maximally uncertain distribution and mutually unbiased basis ($p(s^\dagger|s')=\frac{1}{|\mathcal{S}|}$) allows us to implement thermodynamically efficient erasure.

 While we were able to find the mutually unbiased time-reversed bases for a two-state computation, it's not clear to us how to do this in general.  Thus, we explore different quantum bases for $N$-state computations numerically by sampling using the Haar measure.  For each quantum basis we compare the dissipation to the ambiguity in the time-reversed state (assuming uniform state density $p(s)=1/|\mathcal{S}|$):
\begin{align}
\H[S^\dagger|S] \equiv -\sum_{n,n'}p(s^\dagger_{n'}|s_n)p(s_n) \ln p(s^\dagger_{n'}|s_n) ~.
\end{align}
We call this the \emph{QTR ambiguity} (QTR stands for Quantum Time-Reversal).  The \emph{QTR ambiguity is zero if and only if the memory basis is classical} and time-reversal is a permutation.  This measure effectively quantifies the ``quantumness'' of the time-reversal.  In the case where the QTR ambiguity is zero, we retain the classical result that entropy production diverges for a reliable erasure operation.

We explore different quantum bases for information storage numerically using methods discussed in detail in App. \ref{app: Exploring Bases}.  We randomly sample bases according to the Haar measure \cite{mezzadri2006generate} for systems of varying dimensions $|\mathcal{S}| \in \{2,3,4,20, 100\}$, and calculate both the average entropy production and QTR ambiguity for a collection of samples, as shown in Fig. \ref{fig:Haar_Dissipation}.  We consider erasure computations (see Eq. \ref{eq:Erasure}) with both a very low error rate ($\epsilon = 10^{-26}$), to parallel CMOS logic circuits \cite{shivakumar2002modeling,Riec20_Balancing}, and a high error rate ($\epsilon=10^{-1}$) to address computations where thermal noise is not tightly controlled.

Across all cases, the computational entropy production is high when the time-reversal is classical (zero QTR ambiguity).  Recall that the dissipation must be on the order of $\ln (1/\epsilon)$ in the classical case \cite{boyd2021time}.  As the QTR ambiguity increases, introducing quantum effects, we see decreasing entropy production.  For 2-state systems, we see a monotonic relationship between entropy production and QTR ambiguity, confirming the advantage of computing with a basis with higher ambiguity. This one-to-one relationship is violated for higher dimensions, although the rough trend remains. However, as the dimension increases, the distribution narrows in on a single point where the QTR ambiguity is high and the dissipation is lower than the classical case.  This means that virtually all quantum bases (sampled from the Haar measure) in a large quantum system have the same thermodynamic advantage over classical memory.  Fig. \ref{fig:Haar_Dissipation} shows that the degree of this advantage is tied to the error rate: for high error ($\epsilon= 10^{-1}$) the advantage is on the order of a factor of 2, while the error rates of CMOS circuits ($\epsilon= 10^{-26}$) display orders-of-magnitude thermodynamic quantum advantages.

\section{Conclusion}

Irreversibility in a computation incurs entropy production. This irreversibility depends on both the logical operation as well as the time-reversal symmetries of the underlying memory hardware. We see that quantum hardware allows us to circumnavigate divergent dissipation that occurs in logically irreversible operations.  This advantage arises through the continuum of possible anti-unitary time-reversals in quantum systems, while classical systems can only realize finitely many possible involutions.  Quantum memory fundamentally alters the thermodynamic scaling of logically irreversible computation, replacing classical logarithmic divergence with bounded dissipation.

\paragraph{Acknowledgements}--- We are grateful to Kyle J. Ray, Wesley L. Boyd, and Derek F. Jackson Kimball for insightful conversations and comments.  We recognize the support of the Federation of American Scientists.

\clearpage
\onecolumngrid
\appendix

\section{Random Variable for Thermodynamic Experiments}
\label{app:Random Variables}

Random variables are useful for accurately expressing complex multipartite probabilities and their relationships.  Generally, we will use lower-case letters like $a$ and $b$ for realizations of probability distributions, and upper-case letters like $A$ and $B$ for their random variables, such that we can express marginal probabilities, joint probabilities, and conditionals efficiently
\begin{align}
\text{Marginal probability of }A\text{ realizing }a \text{: }&\Pr(A=a)
\\ \text{Joint probability of }A \text{ and } B\text{ realizing }a \text{ and }b\text{: }&\Pr(A=a,B=b)
\\ \text{Conditional probability of }A \text{ realizing }a \text{ given that } B \text{ realizes }b\text{: }&\Pr(A=a|B=b)\equiv \frac{\Pr(A=a,B=b)}{\Pr(B=b)}.
\end{align}
This also allows us to easily and efficiently express Shannon entropies (measured in nats)
\begin{align}
\H[A] \equiv -\sum_{a \in \mathcal{A}} \Pr(A=a) \ln \Pr(A=a),
\end{align}
and conditional uncertainty
\begin{align}
\H[A|B]\equiv \H[A,B]-\H[B].
\end{align}

\section{Microscopic Reversibility of Hamiltonian Control}
\label{app:Microscopic Reversibility}

Quantum dynamics guarantees specific relationships between the probabilities of forward experiments and reverse experiments.  
Specifically, if we apply Hamiltonian control $H_{\lambda_t}$ to a system $\mathcal{Z}$ (which in our framework is the system--environment supersystem), indexed by a vector of control parameters $\lambda_t$ over the time interval $[0, \tau]$, the net result is a unitary time evolution operator. 
Without loss of generality, we can simplify the analysis by considering a sequence of $N+1$ discrete time-steps of duration $\Delta t= \tau/N$, 
%inducing the particular 
marked by the sequence of 
times $t_n = n \tau/N$,
and take the limit of $N\to\infty$ for continuous-time control.
We then obtain the 
unitary time evolution operator
and its inverse
\begin{align}
U_{\lambda_{0:\tau}}
&=
\lim_{N \to \infty}
\prod_{n=0}^{N} e^{-i \Delta t H_{\lambda_{t_n}}/\hbar}
\\
U_{\lambda_{0:\tau}}^{-1}
&=
\lim_{N \to \infty}
\prod_{n=0}^{N} e^{i \Delta t H_{\lambda_{\tau-t_n}}/\hbar} ~,
\end{align}
where the time ordering of multiplication is implied.

%\begin{align}
%U_{\lambda_{0:\tau}}&=\prod_{t=0}^{\tau} e^{-i t \mathcal{H}_{\lambda_t}/\hbar}
%\\U_{\lambda_{0:\tau}}^{-1}&=\prod_{t=0}^{\tau} e^{i t \mathcal{H}_{\lambda_{\tau-t}}/\hbar}
%\end{align}
We can evaluate the probabilities of measurement outcomes for the final state $z_\tau$ given the initial state was $z_0$ and we perform the experiment with forward control $\lambda_{0:\tau}$:
\begin{align}
\Pr(Z_\tau=z_\tau|Z_0=z_0,\Lambda_{0:\tau}=\lambda_{0:\tau})=|\langle z_\tau| U_{\lambda_{0:\tau}}|z_0 \rangle|^2.
\end{align}
Note that we have explicitly conditioned on the control parameter sequence $\lambda_{0:\tau}$ 
since we consider
%to indicate that we are considering 
distributions from multiple different experiments.  

For comparison, we evaluate the probabilities of transitions under reverse control, where we time reverse the sequence of control parameters, and we conjugate each control parameter
\begin{align}
\lambda^R_{0:\tau} \equiv \lambda_\tau^\dagger \cdots \lambda_0^\dagger.
\end{align}
Time-reversal of the control parameter time-reverses the Hamiltonian $H_{\lambda^\dagger}=\Theta H_{\lambda}\Theta^{-1}$.  This allows us to generalize beyond Hamiltonians that are time-reversal invariant, allowing for the possibility that $\Theta H\neq H\Theta$ \cite{jarzynski2004classical, sakurai2020modern}. In the classical context, time-reversal of the Hamiltonian flips the sign of time-odd parameters like magnetic fields.  However, quantum time-reversal is 
more nuanced.
%less clear.  
%[PMR: What is meant by this last sentence?  Is there something the reader should be cautious about? Also, is there any way to justify $\Theta H_{\lambda}\Theta^{-1}$ beyond the fact that it appears in Sakurai's textbook?]
We can evaluate the probability of measurement outcome $z_0^\dagger$ after applying the control $\lambda^R_{0:\tau}$, given that the initial state was $z_\tau^\dagger$:
\begin{align}
\Pr(Z_\tau=z_0^\dagger|Z_0=z_\tau^\dagger,\Lambda_{0:\tau}=\lambda^R_{0:\tau})&=|\langle z_0^\dagger| U_{\lambda^R_{0:\tau}}|z_\tau^\dagger \rangle|^2.
\end{align}
To evaluate this, the unitary evolution associated with time-reversed control is
\begin{align}
U_{\lambda^R_{0:\tau}} & =
\lim_{N \to \infty}
\prod_{n=0}^{N} e^{-i \Delta t H_{\lambda_{\tau-t_n}^\dagger}/\hbar}
\\ & =
\lim_{N \to \infty}
\prod_{n=0}^{N} e^{-i \Delta t \Theta H_{\lambda_{\tau-t_n}}\Theta^{-1}/\hbar} 
\\ & =
\lim_{N \to \infty}
\prod_{n=0}^{N} e^{\Theta(i \Delta t H_{\lambda_{\tau-t_n}}/\hbar)\Theta^{-1}} .
\end{align}
We can further simplify by noting that the exponential of a time-reverse operator is the time reversal of the exponential of that operator
\begin{align}
    e^{\Theta \hat{O} \Theta^{-1}}& = \sum_{j=0}^\infty \frac{(\Theta \hat{O} \Theta^{-1})^j}{j!}
    \\ & = \Theta\sum_{j=0}^\infty \frac{\hat{O}^j}{j!}\Theta^{-1}
    \\& = \Theta e^{\hat{O}}\Theta^{-1},
\end{align}
such that
\begin{align}
    U_{\lambda^{R}_{0:\tau}}& =
    \lim_{N \to \infty}
    \prod_{n=0}^{N} \Theta e^{i \Delta t H_{\lambda_{\tau-t_n}}/\hbar}\Theta^{-1}
    \\ & = \Theta U_{\lambda_{0:\tau}}^\dagger \Theta^{-1}.
\end{align}
Note that this is more general than the expression $U \Theta =\Theta U^\dagger$ that applies for time-reversal invariant Hamiltonians that obey $H\Theta=\Theta H$ \cite{jarzynski2004classical}.
 Plugging this into the
\begin{align}
\Pr(Z_\tau=z_0^\dagger|Z_0=z_\tau^\dagger,\Lambda_{0:\tau}=\lambda^R_{0:\tau})& =
\lim_{N \to \infty}
|\langle z_0^\dagger|\prod_{n=0}^{N} \Theta e^{i \Delta t H_{\lambda_{\tau-t_n}}/\hbar}\Theta^{-1} |z_\tau^\dagger \rangle|^2
\\ & =|\langle z_0^\dagger|\Theta U^\dagger_{\lambda_{0:\tau}}\Theta^{-1}\Theta |z_\tau \rangle|^2
\\ & =|\langle z_0^\dagger | \Theta U^\dagger_{\lambda_{0:\tau}} |z_\tau \rangle|^2.
\end{align}
Because time reversal is an antiunitary operator $\langle a^\dagger|b^\dagger\rangle=(\langle a | b \rangle)^*$.  Therefore, $|\langle b^\dagger|\Theta|a \rangle|^2=|\langle b| a\rangle |^2$, and we have the condition of microscopic reversibility
\begin{align}
\Pr(Z_\tau=z_0^\dagger|Z_0=z_\tau^\dagger,\Lambda_{0:\tau}=\lambda^R_{0:\tau})&  =|\langle z_0^\dagger | \Theta U^\dagger_{\lambda_{0:\tau}} |z_\tau \rangle|^2
\\ & = |\langle z_\tau|U_{\lambda_{0:\tau}}|z_0 \rangle |^2
\\ & =\Pr(Z_\tau=z_\tau|Z_0=z_0,\Lambda_{0:\tau}=\lambda_{0:\tau})~.
\label{eq:ExplicitMicroRev}
\end{align}
In the main text, we write 
Eq.~\eqref{eq:ExplicitMicroRev}
more succinctly as
\begin{align}
\Pr(z \stackrel{\lambda_{0:\tau}}{\mapsto} z') = 
\Pr(z'^\dagger \stackrel{\lambda_{0:\tau}^R}{\mapsto} z^\dagger)
~.
\end{align}

We can further generalize this idea by constructing forward and reverse experiments that are composed of a trajectory of measurements, 
potentially in different measurement bases and at arbitrary times,
with Hamiltonian control interspersed throughout.

%We redefine $\Lambda_{0:\tau}$ to be the time-series of control parameters including the finite sequence of measurements, and the resulting generalized condition of microscopic reversibility is
%\begin{align}
%\Pr(Z_{0:\tau}=z_{0:\tau}^R|Z_0=z_\tau^\dagger,\Lambda_{0:\tau}=\lambda^R_{0:\tau})=\Pr(Z_{0:\tau}=z_{0:\tau}|Z_0=z_0,\Lambda_{0:\tau}=\lambda_{0:\tau}),
%\label{eq:MeasurementMicroRev}
%\end{align}
%where the time-reversed measurement trajectory is $z^R_{0:\tau}=z_\tau^\dagger \cdots z_0^\dagger$
%with $z_t^R = z_{\tau - t}^\dagger$.
%Moreover, the intervening projective measurements can have arbitrary rank and, in particular,
%can be chosen to be the identity map (i.e., no measurement at all), such that Eq.~\eqref{eq:MeasurementMicroRev} is a strict generalization of Eq.~\eqref{eq:ExplicitMicroRev}.
%[PMR: I'm going to make this part a bit more rigorous, introducing the intervening measurements as I did in my QJarzDFT writeup.]

We now show that 
$U_{\lambda_{t:t'}}^\dagger = \TRinv U_{\lambda_{t:t'}^R} \TR$
and 
$U_{\lambda_{t:t'}} =
\TRinv
U_{\lambda_{t:t'}^R}^\dagger 
\TR$
together
imply a quantum version of microscopic reversibility, applicable to any unitary time evolution interrupted by arbitrary projective measurements at times $\mathcal{T} \subset (0, \tau)$:
%  general measurement protocols:
\begin{align}
\Pr_{\drive, \MSt_{\mathcal{T}}}(Z_\tau = z', \MSt_{\mathcal{T}} = m_1 m_2 \dots m_{|\mathcal{T}|} | Z_0 = z)
= \Pr_{\drive^R, \MSt_{\mathcal{\tilde{T}}}}(Z_\tau = \TR z, \MSt_{\mathcal{\tilde{T}}} = \tilde{m}_{|\mathcal{T}|} \dots \tilde{m}_2 \tilde{m}_1 | Z_0 = \TR z' )  ~,
\end{align}
%
%\begin{align*}
%\Pr_{\drive}(\MSt_{\mathcal{T}} = m_1 m_2 \dots m_N | \MSt_0 = m_1)
%= \Pr_{\smallReverse(\drive)}(\MSt_{\mathcal{\tilde{T}}} = \tilde{m}_N \dots \tilde{m}_2 \tilde{m}_1 | \MSt_0 = \tilde{m}_N )  ~,
%\end{align*}
%
where $\tilde{\mathcal{T}} = (\tau - t)_{t \in \mathcal{T}}$.
%and $\tilde{m}$ indicates detection by the projector $\TR \Pi_m \TRinv$.
The set of projectors 
$\{ \Pi_m\}_m$
used at time $t \in \mathcal{T}$ in the forward protocol are arbitrary, except that they satisfy the typical partitioning of the joint state space:
$\sum_m \Pi_m = I$. 
In particular, the measurement projector $\Pi_m$ has arbitrary rank in an arbitrary basis.
The technical steps parallel the earlier derivation without measurements.

\scalebox{0.76}{
\begin{minipage}{1.3\textwidth}	
\begin{align}
\Pr_{\drive, \MSt_{\mathcal{T}}} & (Z_\tau = z', \MSt_{\mathcal{T}} = m_1 m_2 \dots m_N | Z_0 = z)
\nonumber \\
&= 
 \tr \bigl( \ket{z'} \bra{z'}
 U_{\lambda_{t_{|\mathcal{T}|}:\tau}}
 \Pi_{m_{|\mathcal{T}|}} \dots \Pi_{m_2} 
 U_{\lambda_{t_1:t_2}} \Pi_{m_1} 
 U_{\lambda_{0:t_1}} 
 \ket{z} \bra{z} 
 U_{\lambda_{0:t_1}}^\dagger 
 \Pi_{m_1} 
 U_{\lambda_{t_1:t_2}}^\dagger 
 \Pi_{m_2} \dots \Pi_{m_{|\mathcal{T}|}} 
 U_{\lambda_{t_{|\mathcal{T}|}:\tau}}^\dagger 
 \bigr)
\\
&= 
\tr \bigl( \ket{z'} \bra{z'}
\TRinv
U_{\lambda_{t_{|\mathcal{T}|}:\tau}^R}^\dagger \TR
\Pi_{m_{|\mathcal{T}|}} \dots \Pi_{m_2}
\TRinv
U_{\lambda_{t_1:t_2}^R}^\dagger \TR
\Pi_{m_1} 
\TRinv
U_{\lambda_{0:t_1}^R}^\dagger \TR
\ket{z} \bra{z} 
\TRinv U_{\lambda_{0:t_1}^R} \TR
\Pi_{m_1} 
\TRinv U_{\lambda_{t_1:t_2}^R} \TR
\Pi_{m_2} \dots \Pi_{m_{|\mathcal{T}|}}
\TRinv U_{\lambda_{t_{|\mathcal{T}|}:\tau}^R} \TR
\bigr)
\\
&= 
\tr \bigl(  \TR
\ket{z} \bra{z} 
\TRinv U_{\lambda_{0:t_1}^R} \TR
\Pi_{m_1} 
\TRinv U_{\lambda_{t_1:t_2}^R} \TR
\Pi_{m_2} \dots \Pi_{m_{|\mathcal{T}|}}
\TRinv U_{\lambda_{t_{|\mathcal{T}|}:\tau}^R} \TR
\ket{z'} \bra{z'}
\TRinv
U_{\lambda_{t_{|\mathcal{T}|}:\tau}^R}^\dagger \TR
\Pi_{m_{|\mathcal{T}|}} \dots \Pi_{m_2}
\TRinv
U_{\lambda_{t_1:t_2}^R}^\dagger \TR
\Pi_{m_1} 
\TRinv
U_{\lambda_{0:t_1}^R}^\dagger
\bigr)
\\
&= 
\tr \bigl(  
\TR  \ket{z} \bra{z} \TRinv 
U_{\lambda_{0:t_1}^R} 
\Pi_{\tilde{m}_1} 
U_{\lambda_{t_1:t_2}^R} 
\Pi_{\tilde{m}_2} \dots \Pi_{\tilde{m}_{|\mathcal{T}|}}
U_{\lambda_{t_{|\mathcal{T}|}:\tau}^R} \TR
\ket{z'} \bra{z'}
\TRinv
U_{\lambda_{t_{|\mathcal{T}|}:\tau}^R}^\dagger 
\Pi_{\tilde{m}_{|\mathcal{T}|}} \dots \Pi_{\tilde{m}_2}
U_{\lambda_{t_1:t_2}^R}^\dagger 
\Pi_{\tilde{m}_1} 
U_{\lambda_{0:t_1}^R}^\dagger
\bigr)
\\
&=
\Pr_{\drive^R, \MSt_{\mathcal{\tilde{T}}}}(Z_\tau = \TR z, \MSt_{\mathcal{\tilde{T}}} = \tilde{m}_{|\mathcal{T}|} \dots \tilde{m}_2 \tilde{m}_1 | Z_0 = \TR z' )  ~.
\end{align}
\end{minipage}
}

Note that the time-reversal of measurement outcome $m$ is $\tilde{m}$, detected by the projector $\Pi_{\tilde{m}} = \TR \Pi_m \TRinv$.

\section{A Quantum Detailed Fluctuation Theorem}
\label{app:DFT}

We can apply microscopic reversibility to a joint supersystem of a system $\mathbf{S}$ and environment $\mathbf{E}$ to derive a quantum detailed fluctuation theorem, that relates system trajectory probabilities to entropy production.  
We assume that the
control protocol $\lambda_{0:\tau}$ 
induces Hamiltonian evolution
on the joint system and environment, although the reduced dynamics on the system will in general be non-Hamiltonian.  The parameter $\lambda_t$ specifies the joint Hamiltonian $H_{\lambda_t}$ at time $t$.
%, and therefore implements reversible energy-preserving dynamics. [PMR: That doesn't imply energy conservation.  Consider $H_{\lambda_t} = tI$.] 
Whether through observation of a state sequence in a classical system, or through a sequence of measurements, this drives the system and environment along some stochastic trajectories $s_{0:\tau}$ and $e_{0:\tau}$ with probability:
\begin{align}
F(e_{0:\tau},s_{0:\tau}) \equiv \Pr(E_{0:\tau}=e_{0:\tau},S_{0:\tau}=s_{0:\tau}|\Lambda_{0:\tau}=\lambda_{0:\tau},E_0=e_0,S_0=s_0) ~.
\end{align}
We then consider the probability of reverse trajectories under reverse control $\lambda^R_{0:\tau} \equiv \lambda_\tau^\dagger \cdots \lambda_0^\dagger$:
\begin{align}
R(e_{0:\tau},s_{0:\tau}) \equiv \Pr(E_{0:\tau}=e^R_{0:\tau},S_{0:\tau}=s^R_{0:\tau}|\Lambda_{0:\tau}=\lambda^R_{0:\tau},E_0=e_\tau^\dagger,S_0=s_\tau^\dagger) ~.
\end{align}
As discussed in App.~\ref{app:Microscopic Reversibility}, microscopic reversibility implies that these two probabilities are equivalent:
\begin{align}
R(e_{0:\tau},s_{0:\tau}) =F(e_{0:\tau},s_{0:\tau}).
\end{align}
Furthermore, note that we can multiply by the probability of the environment given the trajectory and system to get the joint probability of the environment trajectory:
\begin{align}
\Pr(E_{0:\tau}=e_{0:\tau},S_{0:\tau}=s_{0:\tau}|\Lambda_{0:\tau}=\lambda_{0:\tau},S_0=s_0)\Pr(E_0=e_0|\Lambda_{0:\tau}=\lambda_{0:\tau},S_0=s_0)^{-1}& =F(e_{0:\tau},s_{0:\tau})
\\ \Pr(E_{0:\tau}=e^R_{0:\tau},S_{0:\tau}=s^R_{0:\tau}|\Lambda_{0:\tau}=\lambda^R_{0:\tau},S_0=s_\tau^\dagger)\Pr(E_0=e^\dagger_\tau|\Lambda_{0:\tau}=\lambda^R_{0:\tau},S_0=s^\dagger_\tau)^{-1}& =R(e_{0:\tau},s_{0:\tau}) ~.
\end{align}
Let us define \emph{surprisal flow} $\phi$  like the entropy flow in the forward control as:
\begin{align}
\label{Eq:SupFlow}
    e^{-\phi(e_0,e_\tau|\lambda_{0:\tau},s_0,s_\tau)} \equiv \frac{\Pr(E_0=e^\dagger_\tau|\Lambda_{0:\tau}=\lambda^R_{0:\tau},S_0=s^\dagger_\tau)}{\Pr(E_0=e_0|\Lambda_{0:\tau}=\lambda_{0:\tau},S_0=s_0)},
\end{align}
which combined with microscopic reversibility implies
\begin{align}
\label{Eq.GeneralDFT}
\Pr(E_{0:\tau}=e_{0:\tau},S_{0:\tau}=s_{0:\tau}|\Lambda_{0:\tau}=\lambda_{0:\tau},S_0=s_0)  e^{-\phi(e_0,e_\tau|\lambda_{0:\tau},s_0,s_\tau)}=\Pr(E_{0:\tau}=e^R_{0:\tau},S_{0:\tau}=s^R_{0:\tau}|\Lambda_{0:\tau}=\lambda^R_{0:\tau},S_0=s_\tau^\dagger).
\end{align}
Note that we can also express the probability of realizing a certain amount of surprisal flow $q$ conditioned on the control, environment, and system trajectories
\begin{align}
\Pr(\Phi = q| E_{0:\tau}=e_{0:\tau},S_{0:\tau}=s_{0:\tau},\Lambda_{0:\tau}=\lambda_{0:\tau})= \delta_{q, \phi (e_0,e_\tau|\lambda_{0:\tau},s_0,s_\tau)}.
\end{align}
The form of Eq.~\eqref{Eq:SupFlow} means that surprisal flow flips under time reversal, such that we can also express the probability of negative entropy flow in the reverse experiment
\begin{align}
\Pr(\Phi = -q| E_{0:\tau}=e^R_{0:\tau},S_{0:\tau}=s^R_{0:\tau},\Lambda_{0:\tau}=\lambda^R_{0:\tau}) & = \delta_{-q, \phi (e^\dagger_\tau,e_0^\dagger|\lambda^R_{0:\tau},s^\dagger_\tau,s_0^\dagger)}
\\ & = \delta_{-q, -\phi (e_0,e_\tau|\lambda_{0:\tau},s,s_\tau)}
\\ & = \delta_{q, \phi (e_0,e_\tau|\lambda_{0:\tau},s,s_\tau)}.
\end{align}
If we multiply both sides of Eq.~\eqref{Eq.GeneralDFT} by this function, we are able to also evaluate the probability of surprisal flows in both cases:
\begin{align}
& \Pr(\Phi=q, E_{0:\tau}=e_{0:\tau},S_{0:\tau}=s_{0:\tau}|\Lambda_{0:\tau}=\lambda_{0:\tau},S_0=s_0)  e^{-\phi (e_0,e_\tau|\lambda_{0:\tau},s_0,s_\tau)}
\\ & =\Pr(\Phi = - q, E_{0:\tau}=e^R_{0:\tau},S_{0:\tau}=s^R_{0:\tau}|\Lambda_{0:\tau}=\lambda^R_{0:\tau},S_0=s_\tau^\dagger).
\end{align}
Note that since this only evaluates to a nonzero number when $q= \phi (e_0,e_\tau|\lambda_{0:\tau},s_0,s_\tau)$, we can rewrite this expression to have the specific entropy flow in the exponential:
\begin{align}
& \Pr(\Phi=q, E_{0:\tau}=e_{0:\tau},S_{0:\tau}=s_{0:\tau}|\Lambda_{0:\tau}=\lambda_{0:\tau},S_0=s_0)  e^{-q}
\\ & =\Pr(\Phi = - q, E_{0:\tau}=e^R_{0:\tau},S_{0:\tau}=s^R_{0:\tau}|\Lambda_{0:\tau}=\lambda^R_{0:\tau},S_0=s_\tau^\dagger).
\end{align}

Finally, we can sum over the possible environmental configurations to find the marginal distribution and obtain something much like Jarzynski's DFT from Hamiltonian equations of motion
\begin{align}
\label{Eq:flowDFT}
\Pr(\Phi=q, S_{0:\tau}=s_{0:\tau}|\Lambda_{0:\tau}=\lambda_{0:\tau},S_0=s_0)  e^{-q} =\Pr(\Phi = - q, S_{0:\tau}=s^R_{0:\tau}|\Lambda_{0:\tau}=\lambda^R_{0:\tau},S_0=s_\tau^\dagger).
\end{align}
This is true in generality, for correlated reservoirs in contact with a system that together can be either classical or quantum.  In addition, we can define a quantity like the change in system entropy, which has been referred to as the ``system surprisal difference'' \cite{wimsatt2022trajectory}:
\begin{align}
e^{\Delta s_\text{sys}(\lambda_{0:L},s_0,s_\tau)}=  \frac{\Pr(S_0=s_0|\Lambda_{0:\tau}=\lambda_{0:\tau})}{\Pr(S_0=s_\tau^\dagger|\Lambda_{0:\tau}=\lambda_{0:\tau}^R)}.
\end{align}

We can then modify Eq. \ref{Eq:flowDFT} to obtain a detailed fluctuation theorem for the total entropy difference $\sigma= \phi+\Delta s_\text{sys}$ :
\begin{align}
\label{Eq:entDFT}
\Pr(\Phi=q, S_{0:\tau}=s_{0:\tau}|\Lambda_{0:\tau}=\lambda_{0:\tau},S_0=s_0)  e^{-q} & =\Pr(\Phi = - q, S_{0:\tau}=s^R_{0:\tau}|\Lambda_{0:\tau}=\lambda^R_{0:\tau},S_0=s_\tau^\dagger)
\\ \frac{\Pr( \Phi=q, S_{0:\tau}=s_{0:\tau}|\Lambda_{0:\tau}=\lambda_{0:\tau})}{\Pr(S_0=s_0|\Lambda_{0:\tau}=\lambda_{0:\tau})}  e^{-q} & =\frac{\Pr(\Phi = - q, S_{0:\tau}=s^R_{0:\tau}|\Lambda_{0:\tau}=\lambda^R_{0:\tau})}{\Pr(S_0=s_\tau^\dagger|\Lambda_{0:\tau}=\lambda^R_{0:\tau})}
\\ \Pr(\Phi=q, S_{0:\tau}=s_{0:\tau}|\Lambda_{0:\tau}=\lambda_{0:\tau})  e^{-q-\Delta s_\text{sys}(\lambda_{0:L},s_0,s_\tau)} & =\Pr(\Phi = - q, S_{0:\tau}=s^R_{0:\tau}|\Lambda_{0:\tau}=\lambda^R_{0:\tau}).
\end{align}
This gives us statistics for calculating the ``entropy difference'' directly from entropy flow and state trajectory statistics that result from two different control protocols $\lambda_{0:\tau}$ and its reverse $\lambda_{0:\tau}^R$.  More concisely, we can express the ``total entropy difference''
\begin{align}
\sigma(q,s_{0:\tau}) & = \ln \frac{f(q,s_{0:\tau})}{r(q,s_{0:\tau})}
\end{align}
where we have defined the forward system process
\begin{align}
f(q,s_{0:\tau}) \equiv  \Pr(\Phi=q, S_{0:\tau}=s_{0:\tau}|\Lambda_{0:\tau}=\lambda_{0:\tau}),
\end{align}
and the reverse system process
\begin{align}
r(q,s_{0:\tau}) \equiv  \Pr(\Phi = - q, S_{0:\tau}=s^R_{0:\tau}|\Lambda_{0:\tau}=\lambda^R_{0:\tau}).
\end{align}

The distributions shown in Eq.~\eqref{Eq:entDFT} would be the raw information one would store by running two experiments: a forward experiment with forward control $\lambda_{0:\tau}$ and reverse control $\lambda_{0:\tau}^R$.  We have not made any assumption of how these two experiments have been initialized, meaning that this equation for the total entropy difference applies in full generality (including initial system-environment correlation, initial environment disequilibrium).  In addition, we obtain a version of the second Law of thermodynamics, where we are guaranteed non-negative average entropy difference by the fact that the average is a relative entropy between forward and reverse experimental distributions 
\begin{align}
\langle \sigma \rangle & = \sum_{q,s_{0:\tau}} \Pr(\Phi=q, S_{0:\tau}=s_{0:\tau}|\Lambda_{0:\tau}=\lambda_{0:\tau}) \ln \frac{\Pr( \Phi=q, S_{0:\tau}=s_{0:\tau}|\Lambda_{0:\tau}=\lambda_{0:\tau})}{\Pr(\Phi = - q, S_{0:\tau}=s^R_{0:\tau}|\Lambda_{0:\tau}=\lambda^R_{0:\tau})}
\\ & = \text{D}_\text{KL}[ f \| r ]~.
\end{align}
Thus we have arrived at mathematical general second Laws of thermodynamics with the only assumption being microscopic reversibility, as would result from Hamiltonian classical or quantum dynamics.

We must insist on the special properties of the initial distributions such that we can interpret this entropy $\Sigma$ physically.  If we make the following assumptions about both the forward and reverse experiments:
\begin{enumerate}
\item The environment's Hamiltonian doesn't depend on the control parameter,
\item The environment is initially in metastable equilibrium $\pi(e)$,
\item The environment equilibrium state probability is preserved under time-reversal $\pi(e) = \pi(e^\dagger)$,
\item The environment's equilibrium doesn't depend on the system state,
\item The initial system distribution of the reverse experiment is the conjugate distribution of the final state of the forward experiment:
\begin{align}
\Pr(S_0=s^\dagger_\tau|\Lambda_{0:\tau}=\lambda_{0:\tau}^R)=\Pr(S_\tau=s_\tau|\Lambda_{0:\tau}=\lambda_{0:\tau}).
\end{align}
\end{enumerate}
then the resulting total entropy difference is the \emph{total entropy production} of the forward experiment.  

The surprisal flow is the change in extensive variables, like the change in particle number $\Delta N_i$ in a reservoir, and heat flow into a reservoir $Q_i$ of where $i$ indexes the $i$th reservoir within the environment:
\begin{align}
\phi (e_0,e_\tau| \lambda_{0:\tau}, s_0, s_\tau) & = - \ln \frac{\pi(e_\tau)}{\pi(e_0)}
\\ &= \sum_{i} \frac{Q_i(e_0 \rightarrow e_\tau)-\mu_i \Delta N_i (e_0 \rightarrow e_\tau)}{\kB T_i}.
\end{align}
The system surprisal difference is the change in system surprisal of the forward experiment
\begin{align}
    \Delta s_\text{sys}(s_0,s_\tau,\lambda_{0:\tau}) & = \ln \frac{\Pr(S_0=s_0|\Lambda_{0:\tau}=\lambda_{0:\tau})}{\Pr(S_\tau=s_\tau|\Lambda_{0:\tau}=\lambda_{0:\tau})}.
\end{align}
With condition 5) as stated above, the surprisal difference multiplied by Boltzmann's constant is the system entropy change that occurs from after the initial measurement to after final measurement 
\begin{align}
   \Delta S_{s_0,s_\tau}&=k_B( \ln \langle s_0 |\rho_0 |s_0 \rangle -\ln \langle s_\tau |\rho_\tau |s_\tau \rangle),
\end{align}
where $\rho_0$ is the state before the initial measurement and $\rho_\tau$ is the state before the final measurement for the forward process.
This entropy change is for a particular pair of measurement outcomes $s_0$ and $s_\tau$, but when averaged over all trajectories, it evaluates to the change in von Neumann entropy of the system.  In this case, the total entropy difference multiplied by Boltzmann's constant is the familiar quantity of entropy production from after the initial measurement to after the final measurement
\begin{align}
\kB \sigma= \Sigma \equiv \Delta S_{s_0,s_\tau}+  \sum_{i} \frac{Q_i-\mu_i \Delta N_i }{ T_i}.
\end{align}
This provides a quantum definition of entropy production of a sequence of measurements that, when averaged, yields the standard definition of quantum entropy \cite{riechers2021initial, Espo10a, deffner2011nonequilibrium}. Note that this explicitly excludes the entropy that would be produced during the initial measurement if the initial state $\rho_0$ is coherent in the measurement basis.  The detailed fluctuation theorem and resulting Second law requires at least the assumption of initially uncorrelated and in equilibrium.

We then consider the contribution to entropy production from a computation that maps $s_0$ to $s_\tau$, rewriting the detailed fluctuation theorem 
\begin{align}
    f(q,s_{0:\tau})e^{-\Sigma(q,s_{0:\tau})/\kB}=r(q,s_{0:\tau}).
\end{align}
We then sum over all intermediate values of the trajectory $s_{0^+:\tau^-}$ and entropy flows $q$ to obtain a relation for the probabilities of start states
\begin{align}
f(s_0,s_\tau)\sum_{q,s_{0^+,\tau^-}}f(q,s_{0^+:\tau^-}|s_0,s_\tau)e^{-\Sigma(q,s_{0:\tau})/\kB}=r(s_0,s_\tau),
\end{align}
where
\begin{align}
    f(s_0,s_\tau)& = \Pr(S_0=s_0,S_\tau=s_\tau|\Lambda_{0:\tau}=\lambda_{0:\tau})
    \\ f(q,s_{0^+:\tau^-}|s_0,s_\tau) & =\Pr(\Phi=q, S_{0^+:\tau^-}=s_{0^+:\tau^-}|S_0=s_0,S_\tau=s_\tau,\Lambda_{0:\tau}=\lambda_{0:\tau})
    \\ r(s_0,s_\tau) &= \Pr(S_0=s_\tau^\dagger,S_\tau=s_0^\dagger|\Lambda_{0:\tau}=\lambda^R_{0:\tau}).
\end{align}
Using Jensen's inequality, we find that the average exponential entropy production is greater than or equal to the exponential of the average entropy production
\begin{align}
\sum_{q,s_{0^+,\tau^-}}f(q,s_{0^+:\tau^-}|s_0,s_\tau)e^{-\Sigma(q,s_{0:\tau})/\kB} \geq e^{-\Sigma_{s_0, s_\tau}/\kB},
\end{align}
where $\Sigma_{s_0,s_\tau}\equiv \sum_{q,s_{0^+,\tau^-}}f(q,s_{0^+:\tau^-}|s_0,s_\tau)\Sigma(q,s_{0:\tau})$ is the average entropy produced when transitioning from $s_0$ to $s_\tau$. Thus, we can bound the average entropy production of a transition from $s_0$ to $s_\tau$ with the forward and reverse transition probabilities
\begin{align}
\Sigma_{s_0,s_\tau} \geq \kB \ln \frac{f(s_0,s_\tau)}{r(s_0,s_\tau)}.
\end{align}
This is the irreversibility of the computation, which we denote the \emph{computational entropy production}, because it bounds the entropy produced by any process that implements this net computation
\begin{align}
    \Sigma^\text{comp}_{s_0,s_\tau} \equiv \kB \ln \frac{f(s_0,s_\tau)}{r(s_0,s_\tau)}.
\end{align}

\section{Binary Quantum Bases}
\label{app: Binary Quantum Bases}

With the spatial time-reversal invariant states $\Theta |x_0 \rangle = |x_0 \rangle $ and $\Theta |x_1 \rangle = |x_1 \rangle$, we can also consider a new 
orthonormal
basis $\{|a \rangle, |b \rangle \}$ which is a superposition of these states
\begin{align}
    |a \rangle \equiv \frac{|x_0 \rangle + i |x_1 \rangle}{\sqrt{2}} \text{, } \; 
    |b \rangle \equiv \frac{|x_0 \rangle - i |x_1 \rangle}{\sqrt{2}}
\end{align}
Upon time reversal, the states swap:
\begin{align}
    \Theta |a \rangle  = |b\rangle
    \text{, } \; 
    \Theta |b \rangle  = |a\rangle.
\end{align}
If we choose the memory to be represented in this basis ($| \zero \rangle = |a  \rangle$ and $| \one \rangle = |b  \rangle$), then we have the familiar time-odd time symmetry described above ($\TR | \zero \rangle = | \one \rangle$ and $\TR | \one \rangle = |\zero \rangle $).  %With these two bases in the same Bloch sphere, we are able to explore the two relevant classical time symmetries.

%\begin{figure}
%\includegraphics[width=.6\columnwidth]{Diagrams/ErasureError.pdf}
%\caption{An erasure operation resets to $0$ with error rate $\epsilon$.}
%\label{fig:ErasureError} 
%\end{figure}

We can explore many more time-reversal symmetries by considering the continuous variety of pure states on the Bloch sphere as shown in Fig.~\ref{fig:BlochReversal}.  These can be parametrized by the angle $\gamma$ from $|a \rangle$ and the 
azimuth $\varphi$ 
%around the 
%Bloch sphere
%$z$-axis 
\begin{align}
|\gamma, \varphi \rangle = \cos (\gamma/2) |a \rangle + e^{i \varphi} \sin (\gamma/2) |b \rangle.
\end{align}
The natural orthogonal state to complete this basis is $|\gamma + \pi, \varphi \rangle$, which is on the opposite side of the Bloch sphere.  Applying the time-reversal operator yields
\begin{align}
   \TR |\gamma, \varphi \rangle  = e^{-i \varphi}|\pi-\gamma, \varphi \rangle.
\end{align}
In essence, the time reversal is a reflection in the Bloch sphere across the plane that corresponds to the angle $\gamma=\pi/2$ as shown in Fig. \ref{fig:BlochReversal}.

To identify the quantum thermodynamic advantage, we consider the angle in the Bloch sphere $\gamma = \pi/4$ and identify the computational states and their time-reversals:
\begin{align}
    | \zero  \rangle  = |\pi/4,\varphi \rangle & \text{, } \;
    | \one  \rangle  = |\pi+\pi/4,\varphi \rangle
    \\ | \zero^\dagger \rangle  = |\pi-\pi/4,\varphi \rangle &\text{, } \;
    | \one^\dagger  \rangle  = |-\pi/4,\varphi \rangle.
\end{align}
These states are mutually unbiased
\begin{align}
\langle \zero^\dagger | \zero  \rangle= \langle \one^\dagger | \zero  \rangle = \langle \zero^\dagger | \one  \rangle= \langle \one^\dagger | \one  \rangle =\frac{1}{\sqrt{2}}.
\end{align}
Choosing $\phi=0$ for one such mutually-unbiased basis, we can express the new information-storing basis in terms of the original positional basis
\begin{align}
| \zero \rangle & = |\pi/4, 0 \rangle 
\\ & = \cos( \pi/8) | a \rangle + \sin( \pi/8) |b \rangle
\\ & = \cos( \pi/8) \frac{|x_0 \rangle + i |x_1 \rangle}{\sqrt{2}}+ \sin( \pi/8) \frac{|x_0 \rangle - i |x_1 \rangle}{\sqrt{2}}
\\ & = \frac{ \cos( \pi/8) + \sin( \pi/8) }{\sqrt{2}}| x_0 \rangle+i\frac{ \cos( \pi/8) - \sin( \pi/8) }{\sqrt{2}}| x_1 \rangle
\\ | \one \rangle & = | 5\pi/4, 0 \rangle 
\\ & = \cos(5 \pi/8) | a \rangle + \sin( 5\pi/8) |b \rangle
\\ & = \cos( 5\pi/8) \frac{|x_0 \rangle + i |x_1 \rangle}{\sqrt{2}}+ \sin( 5\pi/8) \frac{|x_0 \rangle - i |x_1 \rangle}{\sqrt{2}}
\\ & = \frac{ \cos( \pi/2+\pi/8) + \sin( \pi/2 +\pi/8) }{\sqrt{2}}| x_0 \rangle+i\frac{ \cos(\pi/2+ \pi/8) - \sin( \pi/2+\pi/8) }{\sqrt{2}}| x_1 \rangle
\\ & = \frac{ \sin( -\pi/8) + \cos(-\pi/8) }{\sqrt{2}}| x_0 \rangle+i\frac{ \sin(- \pi/8) - \cos( -\pi/8) }{\sqrt{2}}| x_1 \rangle
\\ & = \frac{ -\sin( \pi/8) + \cos(\pi/8) }{\sqrt{2}}| x_0 \rangle+i\frac{ -\sin( \pi/8) - \cos( \pi/8) }{\sqrt{2}}| x_1 \rangle
\\ & = \frac{ \cos( \pi/8) - \sin(\pi/8) }{\sqrt{2}}| x_0 \rangle-i\frac{ \cos( \pi/8) + \sin( \pi/8) }{\sqrt{2}}| x_1 \rangle
\end{align}

\section{Exploring Different Quantum Bases}
\label{app: Exploring Bases}

We explore different bases by considering a basis $\{|n \rangle \}$ which is invariant under time reversal $|n^\dagger \rangle = |n \rangle$, meaning that time-reversal is complex conjugation $\TR = K$ in this basis.  We then sample random unitary $U$ from the Haar measure \cite{mezzadri2006generate}, yielding a new computational basis with elements:
\begin{align}
    |s_n \rangle = U |n \rangle.
\end{align}
In this case, we have the simple expression for the time reversal of each memory state:
\begin{align}
|s_n^\dagger \rangle =U^* |n \rangle.
\end{align}
The resulting overlap between elements of the new basis is
\begin{align}
    \langle s_{n'}^\dagger | s_{n }\rangle =\langle n'|U^\top U |n \rangle.
\end{align}
We can then calculate minimum average entropy production $\langle \Sigma \rangle$ if the erasure uses this computational basis.  The relevance of the time symmetries comes from the probability $p(s^\dagger_{n'}|s_n)\equiv |\langle s_{n'}^\dagger | s_{n }\rangle|^2$ of the time-reversed state $s^\dagger_{n'}$ given the state $s_{n}$.

\clearpage
\twocolumngrid

\newpage
\bibliography{chaos}

\end{document}